\documentclass[aps,prl,twocolumn,showpacs,showkeys]{revtex4-1}
\usepackage{graphicx}% Include figure files
\begin{document}

\title{A microscopic theory for ultra-near-field radiation}
\author{Jian-Sheng Wang}
\author{Jiebin Peng}
%\email[]{phywjs@nus.edu.sg}
%\homepage[]{http://staff.science.nus.edu.sg/~phywjs}
%\thanks{}
\affiliation{Department of Physics, National University of Singapore, Singapore 117551, Republic of Singapore}

\date{10 July 2016}

\begin{abstract}
Using the nonequilibrium Green's function (NEGF) formalism, we propose a microscopic theory for near-field radiative heat transfer between metal plates.  
Tight-binding models for the electrons are coupled to the electromagnetic field continuum. Our approach differs from the established ones based on fluctuational electrodynamics, in that it describes truly nonequilibrium steady states, and is nonlocal in system's dielectric properties.  
For a two quantum-dot model a new length scale emerges at which the heat current shows a peak.  This length scale is related to the physics of parallel plate capacitors.  The three-dimensional model results are consistent with the theory of Polder and van Hove except at very short distances.
\end{abstract}


\maketitle

% introduction
The thermal radiation in a cavity can be well-described by Planck's theory of black-body radiation \cite{Planck_book} - a great achievement of twentieth century physics, which started
the quantum physics revolution.  Two plates at temperatures $T_0$ and $T_1$ will transfer radiative heat at a rate proportional to $T_0^4 - T_1^4$ in the black-body limit, following the
Stefan-Boltzmann law. In the 70s both theoretical \cite{PvH} and 
experimental \cite{hargreaves69,domoto70,Ottens11} work have indicated corrections to the far field prediction when the distances between the plates are comparable to the thermal wavelength of the electromagnetic fields. Near-field effects can be as large as a thousand fold that of the far field results \cite{volokitin07,shen09,Song-rev15}.  

Most recently, due to great progress in technology and precision measurements, much closer proximity is possible, on the scale of nanometers, or near contact.  
Some report near-field enhancements as large as  a million fold  that of black-body values
\cite{Kloppstech2015}, much too large for an explanation, while other experimental results are consistent with the established theory \cite{kimsong2015,song2016}.

Polder and van Hove (PvH) \cite{PvH} were the first to give a quantitative theory of near-field radiation using the Rytov's formulation of fluctuating electromagnetic fields \cite{Rytov,bimonte16}.  The current-current correlation is assumed to follow the fluctuation-dissipation theorem.  The average value of the Poynting vector is computed using the solution of macroscopic Maxwell equations.   In this picture, the near-field contribution is largely due to evanescent modes which are absent in the far field.  
A quantum electrodynamics treatment with linear media and NEGF reproduces the PvH theory \cite{Janowicz03}.

The large near-field effect has also recently been explained by phonon tunneling or surface
phonon polaritons 
\cite{mahanAPL11,xiong14,chiloyan14}.
The aim of this paper is to propose a more fundamental theory. 
In this work, we pay attention to the model of the system that generates the radiation.   Since electrons interact strongly with radiative fields, we begin with
a tight-binding model of the electrons, a metal, for example, and couple it
to the radiative field in a quantized form.  While the electrons are on a discrete lattice, the electromagnetic field is continuous and permeates the whole space.  For simplicity, we ignore the possible role that phonons may play.  We compute
the radiative heat transfer by relating the Green's function of the field to the
normal ordered Poynting vector operator.  The optical property of electrons is
incorporated in the form of photon self-energy from a polarization diagram calculation.  

To illustrate the basic idea, we start with a toy
model consisting of two quantum dots and a one-dimensional (1D) field.  
The same formulation can be applied to more realistic models. 
We report the results of a 3D model calculation with vector potential, the details of which are
presented in the supplemental materials of this paper.

%% section two dots model
We imagine a nanoscale parallel plate capacitor of which the maximum possible charge is
$Q$.  The state of each plate is simplified such that it either has the charge or not.  The plates, located at $z=0$ and $d$, are connected to respective electron baths so that their charges can fluctuate.   The field is taken to be
the scalar potential $\phi(z)$ defined for all $z$.  Photon baths are placed at the far left and right at $-L/2$ and $L/2$, with $L$ much larger than $d$.  The photon bath
is an important feature for a self-consistent and energy conserving theory.  The Hamiltonian of the whole setup, $H = H_{\gamma} + H_e + H_{{\rm int}}$, is 
\begin{eqnarray}
H_\gamma &=& s \int dz\, \frac{1}{2}\left[ \dot{\phi}^2 + c^2 \left({\partial \phi \over \partial z}\right)^2 \right], \\
H_e &=& v_0 c_0^\dagger c_0 + v_1 c_1^\dagger c_1 + {\rm baths\ \&\  couplings},\\
H_{{\rm int}} &=& (-Q) c_0^\dagger c_0 \phi(0) + (-Q) c_1^\dagger c_1 \phi(d),
\end{eqnarray}
where $s = A \epsilon_0/c^2$ is a scale factor to give $H_\gamma$ the dimension of energy;  $A$ is the area of the capacitor, $\epsilon_0$ is the vacuum permittivity, and $c$ is the speed of light.  $c_0$, $c_1$, and their hermitian conjugates are fermionic annihilation and creation operators.  The 
photon field can also be expressed as (in the interaction picture) 
\begin{equation}
\label{eq-phi-qantum}
\phi(z,t) = \sum_{q} \sqrt{{\hbar \over 2 \omega_q s L}}\left( a_q 
e^{i(q z-\omega_q t)} + {\rm h.c.}\right),  
\end{equation}
where $\omega_q = c |q| $  is the photon dispersion 
relation, with wavevector $q = 2\pi k/L$, $k$ an arbitrary integer, 
$a_q$ the bosonic annihilation operator of a photon of mode $q$.
h.c.~stands for the hermitian conjugate of the preceding term.
 
Our task is to compute the energy current between the dots.  From continuity requirements of the field energy, we can establish
an expression for the ``Poynting vector'' as $-\epsilon_0 \dot{\phi} \partial \phi/ \partial z$.
However, to obtain a correct quantum version of the operator, we need to
symmetrize the two factors and also, very importantly, demand normal order
\cite{guidry91} (denoted by the colons here):
\begin{equation}
j = - \frac{\epsilon_0}{2} \left[ :\dot{\phi} {\partial \phi \over  \partial z} : +   
:{\partial \phi \over  \partial z}\dot{\phi}: \right] .
\end{equation}
Normal order dictates that we swap the annihilation operator to the right of the creation operator if that is not already the case.  This removes the zero-point 
motion contribution which otherwise would diverge to infinity. 
We can relate the expectation value of
$j$ to the Green's functions of the photons.  The end effect of the normal order is to take only the positive frequency contribution of the Green's function
(a justification depends on omitting correlations between annihilation-annihilation operators, and similarly creation-creation operators, and will be presented elsewhere).  The average energy current per unit area at location $z$ can be obtained from 
\begin{equation}
\langle j(z) \rangle = 
-\epsilon_0 \int_0^\infty { d\omega \over \pi} \hbar \omega {\rm Re}
{ \partial D^<(\omega, z,z') \over \partial z'}\Big|_{z'=z},
\end{equation}
where $D^<(\omega, z,z') = \int_{-\infty}^{+\infty} D^<(z,t;z',0) e^{i \omega t} dt$
is the frequency domain lesser Green's function for the field $\phi$. 

We evoke the machinery of NEGF \cite{haug96,wang08review,wang14rev,aeberhard12} to calculate the required Green's functions.  First, we define the contour-ordered Green's function as
\begin{equation}
D(z,\tau;z',\tau') = -\frac{i}{\hbar} \bigl\langle T_{\tau} \phi(z,\tau) \phi(z',\tau') \bigr\rangle_{\rm noneq},
\end{equation}
where $\tau$ and $\tau'$ are Keldysh contour times, $T_{\tau}$ is the contour order operator, and the average is over a nonequilibrium steady state.  The operators are in the Heisenberg picture.  Transforming into the interaction picture, and using the standard diagrammatic expansion \cite{bruus04}, we can summarize the result in a contour
ordered Dyson equation, which can be organized as pair of equations in real time, the retarded Dyson equation and the Keldysh equation.  Symbolically, for the 
Keldysh equation,  $D^< = D^r \Pi^<_{\rm tot} D^a$, here $\Pi^<_{\rm tot}$ is a sum of the contributions from the nonlinear interactions at the dots, as well as  the contributions from the photon baths. Due to time translational invariance, the equations become simple in the frequency domain, given as, for the Keldysh equation for our 1D model, 
\begin{equation}
D^<(\omega, z,z') = \sum_j D^r(\omega, z,z_j) \Pi^<_j D^a (\omega,z_j,z')
,
\end{equation}
where the sum is over the set $\{z_j\}=\{-L/2, 0, d, L/2\}$ for
$j=\{L,0,1,R\}$.  The
first and last terms are the left and right photon bath contributions,  
$\Pi^<_L = -2\Omega /(e^{\hbar \omega/(k_B T_L)} - 1)$, with $\Omega
= i sc\omega$, and $T_L$ the temperature of the left photon bath, and similarly for
$\Pi^<_R$.   The photon bath self-energies can be obtained from a 
discrete lattice model, which is essentially the same as the model for phonons \cite{jswpre07}, and then 
taking the limit as the lattice constant goes to zero.
$j=0, 1$ terms are contributions from the quantum dots.
The retarded Green's function satisfies
\begin{eqnarray}
D^{r}(\omega, z,z') &=& D_0^{r}(\omega, z,z') +\nonumber \\
&&\!\!\!\!\!\!  \sum_{j,k=0,1} D_0^r(\omega, z, z_j) \Pi^{r}_{jk}(\omega)
D^{r}(\omega, z_k,z'),
\label{eqDr}
\end{eqnarray}
where $D_0^r(\omega, z,z') = e^{i{\omega\over c}|z-z'|}/(2\Omega)$,  is the free photon retarded Green's function.  The advanced Green's function is obtained
by symmetry, $D^{a}(\omega,z,z') = D^{r}(\omega,z',z)^{*}$.
To make a contact with the usual dyadic Green's function method \cite{Song-rev15}, one can 
turn the Dyson equation into a differential equation by operating with the
inverse of the free Green's function. However, 
due to the discrete nature of the problem, $z_j$ takes only a finite set of values. 
The above equation (\ref{eqDr}) can be solved directly, by choosing a finite
set of values of $\{0,d,\cdots\}$.  It becomes a system of linear equations.  

In addition to the Green's functions of the photons, we also need the Green's functions of the
electrons. A similar Dyson equation for the electrons can be established, with the
Green's function $G$ and electron self-energy $\Sigma$.  The problem is completely specified if these self-energies are known.  However, for interacting systems like the electron-photon interaction $H_{{\rm int}}$, no simple closed 
expression is possible (except
the formal Hedin equations \cite{hedin65}).   For the two-dot model, we present a calculation
with the self-consistent Born approximation (SCBA), also known as random phase
approximation \cite{bruus04}.  In this framework the photon self-energy due to the electron-photon interactions is given, in contour time, as ($j,k=0,1$)
\begin{equation}
\Pi_{jk}(\tau,\tau') = -i\hbar Q^2 G_{jk}(\tau,\tau') G_{kj}(\tau',\tau).
\end{equation}
Since the electrons cannot jump from the left lead to the right lead, we only have
nonzero diagonal terms $\Pi_j \equiv \Pi_{jj}$.  The contour expression can be used to derive
the real-time formulas, e.g., the retarded one in the frequency domain needed for solving the Dyson equation is 
\begin{eqnarray}
\Pi_{jk}^r(\omega) &=& -i\hbar Q^2 \int_{-\infty}^{+\infty} {dE\over 2\pi\hbar}
\Big[ G^r_{jk}(E) G^<_{kj}(E-\hbar \omega) \nonumber \\
&&  + \,
G^<_{jk}(E) G^a_{kj}(E-\hbar\omega) \Big].
\end{eqnarray}
The electron retarded Green's function is given by $G^r_{jj}(E) = 
1/\bigl(E - v_j - \Sigma^r_j(E) - \Sigma^r_{n,j}(E)\bigr)$, where the bath contribution to the self-energy is chosen to follow the Lorentz-Drude model, $\Sigma^r_j(E) =  \frac{1}{2} \Gamma_j/(i + E/E_j)$, where $\Gamma_j$ and $E_j$ are the bath model constants.  The lesser Green's function is given by a Keldysh equation,
$G^<_{jj}(E) = G^r_{jj}(E) \left(\Sigma^{<}_j(E) + \Sigma^{<}_{n,j}(E)\right) G^a_{jj}(E)$.
We refer to the literature for the formulas for the self-energies $\Sigma^{r,<}_{n,j}(E)$ of the electrons arising from the Hartree and Fock diagrams under SCBA \cite{lu2007,zhanglifa2013}.

\begin{figure}
\includegraphics[width=\columnwidth]{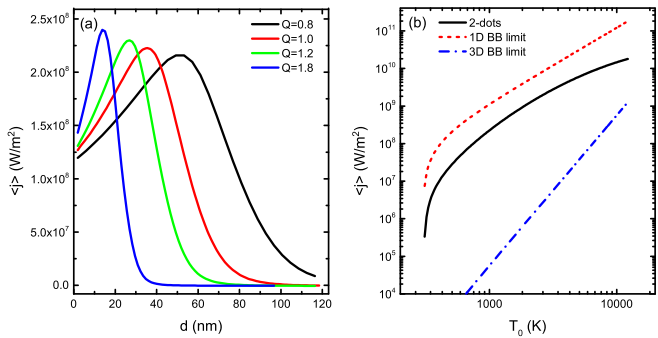}%
\caption{(a) The radiative heat current density $\langle j\rangle$ for the two-dot model as a function of distance $d$ with different maximum charge $Q$. We set the temperatures of the baths at $T_0=1000\,$K, $T_1=300\,$K, $T_L=100\,$K, $T_R=30\,$K, chemical potentials $\mu_0=0\,$eV, $\mu_1=0.02\,$eV, and 
onsite $v_0=0$, $v_1=0.01\,$eV, area $A=389.4$ (nm)$^2$,  and the electron bath parameters $\Gamma_0 = 1\,$eV, $\Gamma_1=0.5\,$eV,
$E_0 =2.0\,$eV, $E_1 = 1.0\,$eV.
(b) The temperature dependence of the radiative heat current density. Here, we set $T_1=300\,$K and vary $T_0$, with $d=35.5\,$nm and $Q= 1\, e$. Other parameters are the same as for Fig.~1(a).}
\end{figure}
%% discuss results
We now discuss the numerical results of the two-dot model.
Figure 1(a) shows the radiative heat current density $\langle j \rangle$ as a function of the two-dot separation $d$ with various charge $Q$. A peak is discovered, and the peak position varies with the parameter $Q$ (or area $A$ not shown). This phenomenon is different from near-field radiative heat transfer results dominated by evanescent modes.  There are no 
evanescent modes in our model.  The results can be understood from the following considerations:  (i) two dots attain the condition of quantum resonance for the 
photon field at a specific distance; (ii) due to the strong coupling between surface charges and photons, radiative heat transfer is enhanced.   The peak position $d_{\rm max}$ is in good
agreement with a parallel plate capacitor model when the energy of the capacitor is of the order of an eV, i.e., $U = Q^2/(2C) \sim 1\,$eV, where $C 
= \epsilon_0 A/d_{\rm max}$ is the capacitance.  More precisely, a length scale can be
obtained by analyzing the expressions of the photon Green's functions, giving
$d_{{\rm max}} = - \frac{1}{2} \epsilon_0 A [1/\Pi^r_0(0)+ 1/\Pi^r_1(0)]$,
where the photon retarded self-energies at the dots are evaluated at zero frequency.
This prediction is closely followed by the data. 

Besides, Figure 1(a) also shows a large value of radiative heat transfer due to  surface charge resonance which is not obtained in the standard fluctuational electrodynamics.  At the peak positions, the radiative heat current density is approximately $2.5 \times 10^8$ W/m$^2$, which is almost five thousand times larger than the black-body limit of $5.6\times 10^4$ W/m$^2$. Compared to a one-dimensional Landauer formula (1D BB) result with perfect transmission, i.e. $1.1\times 10^9$ W/m$^2$, our numbers are about a quarter of that upper limit. 
Such enhancement is mainly due to transverse confinement (there is
only one transmission mode) and the small area $A$.
The temperature dependence of the current density is plotted in Fig.~1(b). 
Asymptotically for large $T_0$ fixing $T_1$,
the Stefan-Boltzmann law gives the fourth power of $T_0$ and 1D BB limit gives a quadratic function of $T_0$.  The quantum dot model demonstrates
an unusual temperature dependence which could be related to the specific density of states of the quantum dots as comparing to bulk systems.

%% 3D model.
We then define the 3D model and discuss its predictions.  We consider
a semi-infinite cubic lattice of lattice spacing $a$ and a cross section of $L^2$ electron sites, with periodic
boundary conditions in the transverse directions.  The layers 1, 2,
to $L_z$ form the right system, and the rest of the sites, $L_z+1$, $L_z+2$, $\cdots$, form the right bath.  The left is similar with the system layers numbered
$-L_z +1$, $\cdots$, $-1$, 0.  The distinction between system and bath is that the baths do not interact with the electromagnetic field. The two semi-infinite blocks are separated by a distance $d$. We take a nearest neighbor hopping model with a hopping
parameter $t$. Only the surface layers numbered 0 
and 1 have onsite potentials, namely $v_0$ and $v_1$.  These potentials mimic the Coulomb interactions of the charges, and will be determined according to a capacitor model, i.e.,
$v_0$ and $v_1$ will be adjusted such that the surface charge per unit area on the plates satisfies $\sigma = \epsilon_0 V/d$, where $V$ is the potential drop across the gap.

Gauge invariance uniquely determines the form of interactions between the electrons and the fields.  Using the Coulomb gauge ($\nabla \cdot {\bf A} = 0$) \cite{mahan00}, the electron system and interaction term can be written as 
\begin{equation}
\label{eqegint}
H_e + H_{\rm int} = \sum_{l,l'} c^\dagger_l H_{ll'} c_{l'} \exp\left( - \frac{ie}{\hbar} 
\int_{l'}^l {\bf A} \cdot d{\bf l}\right),
\end{equation}
where $l=(l_x, l_y, l_z)$ denotes the electron sites, $H_{ll'}$ is the single electron
Hamiltonian matrix element, and ${\bf A}$ is the vector potential.  We have used the convention
that the charge of the electron is $-e$.  The radiative field itself has the Hamiltonian
$H_{\gamma} = \int d^3 {\bf r}\, \frac{1}{2} \left( \epsilon_0 {\bf E}_{\perp}^2 
+ \frac{1}{\mu_0} {\bf B}^2 \right)$, where ${\bf E}_{\perp} = - \partial
{\bf A}/{\partial t}$ and ${\bf B} = \mu_0 {\bf H} =  \nabla \times {\bf A}$. 
 
The principles described earlier apply equally well to this model.  For example, the Poynting vector operator can be defined by
\begin{equation}
{\bf S} = \frac{1}{2} \Bigl(  : {\bf E}_\perp \times {\bf H} : - 
: {\bf H} \times {\bf E}_\perp : \Bigr),
\end{equation}
and the central object for calculations is the contour-ordered photon Green's function
\begin{equation}
D^{\alpha\beta}({\bf r}, \tau; {\bf r}', \tau') = -\frac{i}{\hbar}
\bigl\langle  T_{\tau} A^\alpha({\bf r}, \tau) A^\beta( {\bf r}', \tau') \bigr\rangle.
\end{equation}
The Dyson equation has the same form if we interpret $\Pi^r, D^r_0$, and $D^r$
as $3\times 3$ matrices with 
``spin'' indices, $\alpha,\beta = x,y,z$. 

The supplemental materials present the formulas involved for this calculation. Due to lattice periodicity in the transverse directions, a great
simplification can be made by working in the wave-vector ${\bf q}_\perp$ space.
This essentially reduces it to a 1D problem for each ${\bf q}_\perp$. 
Still, self-consistency is computationally very demanding. Our results below
are based on Born approximations of the self energies.  That is, when computing
the photon self-energy, we use the unperturbed electron Green's function $G_0$.

\begin{figure}
\includegraphics[width=\columnwidth]{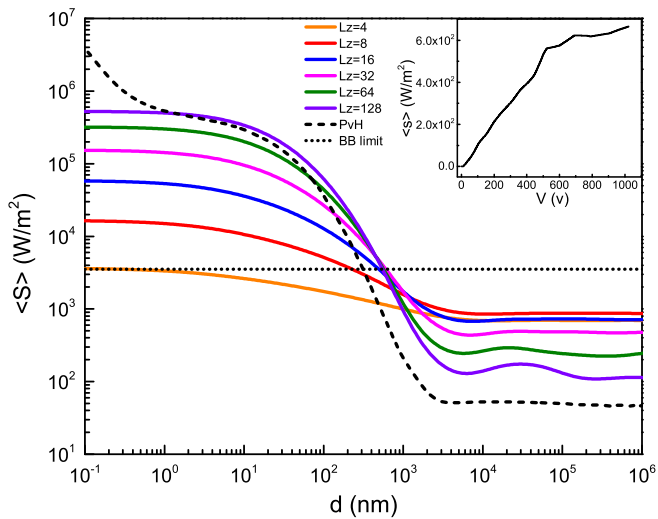}%
\caption{\label{figscan-d} The average Poynting vector $\langle S \rangle$ as a 
function of gap distance $d$ for the 3D cubic lattice model.  We set
the temperatures of electron baths at $T_0 = 500\,$K, $T_1 = 100\,$K,  
$\mu_0 = \mu_1 = v_0 = v_1 = 0\,$eV,
with lattice constant $a=2.117\,$\AA, hopping parameter $t=0.85\,$eV,
damping parameter $\eta = 0.0272\,$eV, and transverse dimension $L=40$.  The 
dashed curve shows the PvH result.  The dotted horizontal line is the black-body limit. The inset shows
$\langle S \rangle$ against voltage bias $V$, when the chemical potentials are set at
$\mu_0 = -6t$, $\mu_1 = 6t$, at a distance $d=5.29\,$nm and $L_z = 4$.}
\end{figure}

%% 3D results
Figure 2 shows the 3D results of the average Poynting vector
$\langle S \rangle$ at the gap of the cubic lattices for different thickness $L_z$,
as well as a PvH calculation 
using a simple Drude model for comparison (with dielectric function given by 
$\epsilon(\omega) = 1 - n e^2/[\epsilon_0 m \omega(\omega + i 2\eta/\hbar)]$, 
where $n=\frac{1}{6}a^{-3}$ to match the conductivity of the model).  
The parameters are chosen to be close to that of a typical metal. 
Our microscopic theory results more or less match the PvH results.  
An exact agreement is not possible since the Drude model is not exact and 
our calculation is limited to finite sizes of $L_z$.  An interesting phenomenon is that the far-field values are dominated
by the first few layers, while near-field values depend on more layers.  We expect
the near-field values to be saturated for sufficiently large $L_z$, although
PvH theory shows a $1/d^2$ divergence.  The
inset shows a dependence of the thermal energy flux on voltage. 
We set the chemical potential $\mu_0$ of bath 0 to the bottom of the band and
$\mu_1$ of bath 1 to the top of the band so that the system behaves like
semiconductors of N-type and P-type, respectively.  
This graph demonstrates that it is possible to change the radiative heat current by applying a voltage bias (although huge bias is required). 

%%% summary part and perspective
In summary, we have presented a more fundamental theory of near-field
radiative heat transfer that applies to distances approaching atomic lattice constants.  From the perspective of this current theory, the limitations of the PvH theory as an approximate treatment are apparent.  First, we note that the current-current correlation is not identical to the photon self-energies; this is only
true to leading order in the diagrammatic expansion.  Second, the use of the
fluctuation-dissipation theorem for the current correlation is consistent with a Born approximation.  A more rigorous treatment such as SCBA will lead to a true 
steady state.  Only the electron baths have well-defined temperatures, and with the electron-photon interaction being nonlinear, we can not recover a 
Landauer-like formula like in the PvH theory
\cite{Biehs10}.  However, our theory does recover the
PvH result if the electrons are treated as an effective medium with translational
invariance in all directions. Specifically, the relation between the retarded self-energy and the dielectric function is $\Pi^{r,\alpha\beta}_{jk}(\omega) = - \epsilon_0 \omega^2 a^3 \left[ \epsilon(\omega) - 1\right] \delta_{\alpha\beta} \delta_{jk}$. 

The approach proposed here opens the way for the treatment of other geometries such a surface and a tip, and ionic phonon systems, or more exotic
systems such as topological insulators or Weyl semi-metals, where surface states may play an important role and greater near-field effects may be present.  Our approach can also be interfaced with first principle calculations, thus enabling more rigorous predictions of near-field properties.

The authors thank Lifa Zhang for stimulating discussions and Han Hoe Yap for pointing out an error in an earlier version of the supplemental materials. This work is supported by FRC grant R-144-000-343-112. 

\bibliography{nearfield}

%merlin.mbs 2010-03-15 4.21a (PWD, AO, DPC)
\begin{thebibliography}{29}%
\makeatletter
\providecommand \@ifxundefined [1]{%
 \@ifx{#1\undefined}
}%
\providecommand \@ifnum [1]{%
 \ifnum #1\expandafter \@firstoftwo
 \else \expandafter \@secondoftwo
 \fi
}%
\providecommand \@ifx [1]{%
 \ifx #1\expandafter \@firstoftwo
 \else \expandafter \@secondoftwo
 \fi
}%
\providecommand \natexlab [1]{#1}%
\providecommand \enquote  [1]{``#1''}%
\providecommand \bibnamefont  [1]{#1}%
\providecommand \bibfnamefont [1]{#1}%
\providecommand \citenamefont [1]{#1}%
\providecommand \href@noop [0]{\@secondoftwo}%
\providecommand \href [0]{\begingroup \@sanitize@url \@href}%
\providecommand \@href[1]{\@@startlink{#1}\@@href}%
\providecommand \@@href[1]{\endgroup#1\@@endlink}%
\providecommand \@sanitize@url [0]{\catcode `\\12\catcode `\$12\catcode
  `\&12\catcode `\#12\catcode `\^12\catcode `\_12\catcode `\%12\relax}%
\providecommand \@@startlink[1]{}%
\providecommand \@@endlink[0]{}%
\providecommand \url  [0]{\begingroup\@sanitize@url \@url }%
\providecommand \@url [1]{\endgroup\@href {#1}{\urlprefix }}%
\providecommand \urlprefix  [0]{URL }%
\providecommand \Eprint [0]{\href }%
\@ifxundefined \urlstyle {%
  \providecommand \doi  [0]{\begingroup \@sanitize@url \@doi}%
  \providecommand \@doi [1]{\endgroup \@@startlink {\doibase
  #1}doi:\discretionary {}{}{}#1\@@endlink }%
}{%
  \providecommand \doi  [0]{doi:\discretionary{}{}{}\begingroup
  \urlstyle{rm}\Url }%
}%
\providecommand \doibase [0]{http://dx.doi.org/}%
\providecommand \Doi [0]{\begingroup \@sanitize@url \@Doi }%
\providecommand \@Doi  [1]{\endgroup\@@startlink{\doibase#1}\@@Doi}%
\providecommand \@@Doi [1]{#1\@@endlink}%
\providecommand \selectlanguage [0]{\@gobble}%
\providecommand \bibinfo  [0]{\@secondoftwo}%
\providecommand \bibfield  [0]{\@secondoftwo}%
\providecommand \translation [1]{[#1]}%
\providecommand \BibitemOpen [0]{}%
\providecommand \bibitemStop [0]{}%
\providecommand \bibitemNoStop [0]{.\EOS\space}%
\providecommand \EOS [0]{\spacefactor3000\relax}%
\providecommand \BibitemShut  [1]{\csname bibitem#1\endcsname}%
%</preamble>
\bibitem [{\citenamefont {Planck}(1906)}]{Planck_book}%
  \BibitemOpen
  \bibfield  {author} {\bibinfo {author} {\bibfnamefont {M.}~\bibnamefont
  {Planck}},\ }\href@noop {} {\emph {\bibinfo {title} {Theorie der
  W\"armestrahlung}}}\ (\bibinfo  {publisher} {J. A. Barth},\ \bibinfo
  {address} {Leipzig},\ \bibinfo {year} {1906})\ \bibinfo {note} {translated
  into English by Morton Masius in M. Planck, \textit{The Theory of Heat
  Radiation}, (Dover, New York, 1991)}\BibitemShut {NoStop}%
\bibitem [{\citenamefont {Polder}\ and\ \citenamefont {{M. Van
  Hove}}(1971)}]{PvH}%
  \BibitemOpen
  \bibfield  {author} {\bibinfo {author} {\bibfnamefont {D.}~\bibnamefont
  {Polder}}\ and\ \bibinfo {author} {\bibnamefont {{M. Van Hove}}},\
  }\href@noop {} {\bibfield  {journal} {\bibinfo  {journal} {Phys. Rev. B},\
  }\textbf {\bibinfo {volume} {4}},\ \bibinfo {pages} {3303} (\bibinfo {year}
  {1971})}\BibitemShut {NoStop}%
\bibitem [{\citenamefont {Hargreaves}(1969)}]{hargreaves69}%
  \BibitemOpen
  \bibfield  {author} {\bibinfo {author} {\bibfnamefont {C.~M.}\ \bibnamefont
  {Hargreaves}},\ }\href@noop {} {\bibfield  {journal} {\bibinfo  {journal}
  {Phys. Lett.},\ }\textbf {\bibinfo {volume} {30A}},\ \bibinfo {pages} {491}
  (\bibinfo {year} {1969})}\BibitemShut {NoStop}%
\bibitem [{\citenamefont {Domoto}\ \emph {et~al.}(1970)\citenamefont {Domoto},
  \citenamefont {Boehm},\ and\ \citenamefont {Tien}}]{domoto70}%
  \BibitemOpen
  \bibfield  {author} {\bibinfo {author} {\bibfnamefont {G.~A.}\ \bibnamefont
  {Domoto}}, \bibinfo {author} {\bibfnamefont {R.~F.}\ \bibnamefont {Boehm}}, \
  and\ \bibinfo {author} {\bibfnamefont {C.~L.}\ \bibnamefont {Tien}},\
  }\href@noop {} {\bibfield  {journal} {\bibinfo  {journal} {J. Heat Transf.},\
  }\textbf {\bibinfo {volume} {92}},\ \bibinfo {pages} {412} (\bibinfo {year}
  {1970})}\BibitemShut {NoStop}%
\bibitem [{\citenamefont {Ottens}\ \emph {et~al.}(2011)\citenamefont {Ottens},
  \citenamefont {Quetschke}, \citenamefont {Wise}, \citenamefont {Alemi},
  \citenamefont {Lundock}, \citenamefont {Mueller}, \citenamefont {Reitze},
  \citenamefont {Tanner},\ and\ \citenamefont {Whiting}}]{Ottens11}%
  \BibitemOpen
  \bibfield  {author} {\bibinfo {author} {\bibfnamefont {R.~S.}\ \bibnamefont
  {Ottens}}, \bibinfo {author} {\bibfnamefont {V.}~\bibnamefont {Quetschke}},
  \bibinfo {author} {\bibfnamefont {S.}~\bibnamefont {Wise}}, \bibinfo {author}
  {\bibfnamefont {A.~A.}\ \bibnamefont {Alemi}}, \bibinfo {author}
  {\bibfnamefont {R.}~\bibnamefont {Lundock}}, \bibinfo {author} {\bibfnamefont
  {G.}~\bibnamefont {Mueller}}, \bibinfo {author} {\bibfnamefont {D.~H.}\
  \bibnamefont {Reitze}}, \bibinfo {author} {\bibfnamefont {D.~B.}\
  \bibnamefont {Tanner}}, \ and\ \bibinfo {author} {\bibfnamefont {B.~F.}\
  \bibnamefont {Whiting}},\ }\href@noop {} {\bibfield  {journal} {\bibinfo
  {journal} {Phys. Rev. Lett.},\ }\textbf {\bibinfo {volume} {107}},\ \bibinfo
  {pages} {014301} (\bibinfo {year} {2011})}\BibitemShut {NoStop}%
\bibitem [{\citenamefont {Volokitin}\ and\ \citenamefont
  {Persson}(2007)}]{volokitin07}%
  \BibitemOpen
  \bibfield  {author} {\bibinfo {author} {\bibfnamefont {A.}~\bibnamefont
  {Volokitin}}\ and\ \bibinfo {author} {\bibfnamefont {B.}~\bibnamefont
  {Persson}},\ }\href@noop {} {\bibfield  {journal} {\bibinfo  {journal} {Rev.
  Mod. Phys.},\ }\textbf {\bibinfo {volume} {79}},\ \bibinfo {pages} {1291}
  (\bibinfo {year} {2007})}\BibitemShut {NoStop}%
\bibitem [{\citenamefont {Shen}\ \emph {et~al.}(2009)\citenamefont {Shen},
  \citenamefont {Narayanaswamy},\ and\ \citenamefont {Chen}}]{shen09}%
  \BibitemOpen
  \bibfield  {author} {\bibinfo {author} {\bibfnamefont {S.}~\bibnamefont
  {Shen}}, \bibinfo {author} {\bibfnamefont {A.}~\bibnamefont {Narayanaswamy}},
  \ and\ \bibinfo {author} {\bibfnamefont {G.}~\bibnamefont {Chen}},\
  }\href@noop {} {\bibfield  {journal} {\bibinfo  {journal} {Nano Lett.},\
  }\textbf {\bibinfo {volume} {9}},\ \bibinfo {pages} {2909} (\bibinfo {year}
  {2009})}\BibitemShut {NoStop}%
\bibitem [{\citenamefont {Song}\ \emph {et~al.}(2015)\citenamefont {Song},
  \citenamefont {Fiorino}, \citenamefont {Meyhofer},\ and\ \citenamefont
  {Reddy}}]{Song-rev15}%
  \BibitemOpen
  \bibfield  {author} {\bibinfo {author} {\bibfnamefont {B.}~\bibnamefont
  {Song}}, \bibinfo {author} {\bibfnamefont {A.}~\bibnamefont {Fiorino}},
  \bibinfo {author} {\bibfnamefont {E.}~\bibnamefont {Meyhofer}}, \ and\
  \bibinfo {author} {\bibfnamefont {P.}~\bibnamefont {Reddy}},\ }\href@noop {}
  {\bibfield  {journal} {\bibinfo  {journal} {AIP Advances},\ }\textbf
  {\bibinfo {volume} {5}},\ \bibinfo {pages} {053503} (\bibinfo {year}
  {2015})}\BibitemShut {NoStop}%
\bibitem [{\citenamefont {Kloppstech}\ \emph {et~al.}()\citenamefont
  {Kloppstech}, \citenamefont {{K\"onne}}, \citenamefont {Biehs}, \citenamefont
  {Rodriguez}, \citenamefont {Worbes}, \citenamefont {Hellmann},\ and\
  \citenamefont {Kittel}}]{Kloppstech2015}%
  \BibitemOpen
  \bibfield  {author} {\bibinfo {author} {\bibfnamefont {K.}~\bibnamefont
  {Kloppstech}}, \bibinfo {author} {\bibfnamefont {N.}~\bibnamefont
  {{K\"onne}}}, \bibinfo {author} {\bibfnamefont {S.-A.}\ \bibnamefont
  {Biehs}}, \bibinfo {author} {\bibfnamefont {A.~W.}\ \bibnamefont
  {Rodriguez}}, \bibinfo {author} {\bibfnamefont {L.}~\bibnamefont {Worbes}},
  \bibinfo {author} {\bibfnamefont {D.}~\bibnamefont {Hellmann}}, \ and\
  \bibinfo {author} {\bibfnamefont {A.}~\bibnamefont {Kittel}},\ }\href@noop {}
  {}\bibinfo {note} {{arXiv}:1510.06311}\BibitemShut {NoStop}%
\bibitem [{\citenamefont {Kim}\ \emph {et~al.}(2015)\citenamefont {Kim},
  \citenamefont {Song}, \citenamefont {{Fern\'andez-Hurtado}}, \citenamefont
  {Lee}, \citenamefont {Jeong}, \citenamefont {Cui}, \citenamefont {Thompson},
  \citenamefont {Feist}, \citenamefont {Reid}, \citenamefont
  {{Garc\'ia-Vidal}}, \citenamefont {Cuevas}, \citenamefont {Meyhofer},\ and\
  \citenamefont {Reddy}}]{kimsong2015}%
  \BibitemOpen
  \bibfield  {author} {\bibinfo {author} {\bibfnamefont {K.}~\bibnamefont
  {Kim}}, \bibinfo {author} {\bibfnamefont {B.}~\bibnamefont {Song}}, \bibinfo
  {author} {\bibfnamefont {V.}~\bibnamefont {{Fern\'andez-Hurtado}}}, \bibinfo
  {author} {\bibfnamefont {W.}~\bibnamefont {Lee}}, \bibinfo {author}
  {\bibfnamefont {W.}~\bibnamefont {Jeong}}, \bibinfo {author} {\bibfnamefont
  {L.}~\bibnamefont {Cui}}, \bibinfo {author} {\bibfnamefont {D.}~\bibnamefont
  {Thompson}}, \bibinfo {author} {\bibfnamefont {J.}~\bibnamefont {Feist}},
  \bibinfo {author} {\bibfnamefont {M.~T.~H.}\ \bibnamefont {Reid}}, \bibinfo
  {author} {\bibfnamefont {F.~J.}\ \bibnamefont {{Garc\'ia-Vidal}}}, \bibinfo
  {author} {\bibfnamefont {J.~C.}\ \bibnamefont {Cuevas}}, \bibinfo {author}
  {\bibfnamefont {E.}~\bibnamefont {Meyhofer}}, \ and\ \bibinfo {author}
  {\bibfnamefont {P.}~\bibnamefont {Reddy}},\ }\href@noop {} {\bibfield
  {journal} {\bibinfo  {journal} {Nature},\ }\textbf {\bibinfo {volume}
  {528}},\ \bibinfo {pages} {387} (\bibinfo {year} {2015})}\BibitemShut
  {NoStop}%
\bibitem [{\citenamefont {Song}\ \emph {et~al.}()\citenamefont {Song},
  \citenamefont {Thompson}, \citenamefont {Fiorino}, \citenamefont {Ganjeh},\
  and\ \citenamefont {Reddy}}]{song2016}%
  \BibitemOpen
  \bibfield  {author} {\bibinfo {author} {\bibfnamefont {B.}~\bibnamefont
  {Song}}, \bibinfo {author} {\bibfnamefont {D.}~\bibnamefont {Thompson}},
  \bibinfo {author} {\bibfnamefont {A.}~\bibnamefont {Fiorino}}, \bibinfo
  {author} {\bibfnamefont {Y.}~\bibnamefont {Ganjeh}}, \ and\ \bibinfo {author}
  {\bibfnamefont {P.}~\bibnamefont {Reddy}},\ }\href@noop {} {\bibfield
  {journal} {\bibinfo  {journal} {Nature Nanotechnology}},\ }\bibinfo {note}
  {dOI:10.1038/NNANO.2016.17}\BibitemShut {NoStop}%
\bibitem [{\citenamefont {Rytov}(1953)}]{Rytov}%
  \BibitemOpen
  \bibfield  {author} {\bibinfo {author} {\bibfnamefont {S.~M.}\ \bibnamefont
  {Rytov}},\ }\href@noop {} {\emph {\bibinfo {title} {Theory of Electric
  Fluctuations and Thermal Radiation}}}\ (\bibinfo  {publisher} {Air Force
  Cambridge Research Center},\ \bibinfo {address} {Bedford, MA},\ \bibinfo
  {year} {1953})\BibitemShut {NoStop}%
\bibitem [{\citenamefont {Bimonte}\ \emph {et~al.}()\citenamefont {Bimonte},
  \citenamefont {Emig}, \citenamefont {Kardar},\ and\ \citenamefont
  {{Kr\"uger}}}]{bimonte16}%
  \BibitemOpen
  \bibfield  {author} {\bibinfo {author} {\bibfnamefont {G.}~\bibnamefont
  {Bimonte}}, \bibinfo {author} {\bibfnamefont {T.}~\bibnamefont {Emig}},
  \bibinfo {author} {\bibfnamefont {M.}~\bibnamefont {Kardar}}, \ and\ \bibinfo
  {author} {\bibfnamefont {M.}~\bibnamefont {{Kr\"uger}}},\ }\href@noop {}
  {}\bibinfo {note} {{arXiv}:1606.03740}\BibitemShut {NoStop}%
\bibitem [{\citenamefont {Janowicz}\ \emph {et~al.}(2003)\citenamefont
  {Janowicz}, \citenamefont {Reddig},\ and\ \citenamefont
  {Holthaus}}]{Janowicz03}%
  \BibitemOpen
  \bibfield  {author} {\bibinfo {author} {\bibfnamefont {M.}~\bibnamefont
  {Janowicz}}, \bibinfo {author} {\bibfnamefont {D.}~\bibnamefont {Reddig}}, \
  and\ \bibinfo {author} {\bibfnamefont {M.}~\bibnamefont {Holthaus}},\
  }\href@noop {} {\bibfield  {journal} {\bibinfo  {journal} {Phys. Rev. A},\
  }\textbf {\bibinfo {volume} {68}},\ \bibinfo {pages} {043823} (\bibinfo
  {year} {2003})}\BibitemShut {NoStop}%
\bibitem [{\citenamefont {Mahan}(2011)}]{mahanAPL11}%
  \BibitemOpen
  \bibfield  {author} {\bibinfo {author} {\bibfnamefont {G.~D.}\ \bibnamefont
  {Mahan}},\ }\href@noop {} {\bibfield  {journal} {\bibinfo  {journal} {Appl.
  Phys. Lett.},\ }\textbf {\bibinfo {volume} {98}},\ \bibinfo {pages} {132106}
  (\bibinfo {year} {2011})}\BibitemShut {NoStop}%
\bibitem [{\citenamefont {Xiong}\ \emph {et~al.}(2014)\citenamefont {Xiong},
  \citenamefont {Yang}, \citenamefont {Kosevich}, \citenamefont {Chalopin},
  \citenamefont {{D' Agosta}}, \citenamefont {Cortona},\ and\ \citenamefont
  {Volz}}]{xiong14}%
  \BibitemOpen
  \bibfield  {author} {\bibinfo {author} {\bibfnamefont {S.}~\bibnamefont
  {Xiong}}, \bibinfo {author} {\bibfnamefont {K.}~\bibnamefont {Yang}},
  \bibinfo {author} {\bibfnamefont {Y.~A.}\ \bibnamefont {Kosevich}}, \bibinfo
  {author} {\bibfnamefont {Y.}~\bibnamefont {Chalopin}}, \bibinfo {author}
  {\bibfnamefont {R.}~\bibnamefont {{D' Agosta}}}, \bibinfo {author}
  {\bibfnamefont {P.}~\bibnamefont {Cortona}}, \ and\ \bibinfo {author}
  {\bibfnamefont {S.}~\bibnamefont {Volz}},\ }\href@noop {} {\bibfield
  {journal} {\bibinfo  {journal} {Phys. Rev. Lett.},\ }\textbf {\bibinfo
  {volume} {112}},\ \bibinfo {pages} {114301} (\bibinfo {year}
  {2014})}\BibitemShut {NoStop}%
\bibitem [{\citenamefont {Chiloyan}\ \emph {et~al.}(2015)\citenamefont
  {Chiloyan}, \citenamefont {Garg}, \citenamefont {Esfarjani},\ and\
  \citenamefont {Chen}}]{chiloyan14}%
  \BibitemOpen
  \bibfield  {author} {\bibinfo {author} {\bibfnamefont {V.}~\bibnamefont
  {Chiloyan}}, \bibinfo {author} {\bibfnamefont {J.}~\bibnamefont {Garg}},
  \bibinfo {author} {\bibfnamefont {K.}~\bibnamefont {Esfarjani}}, \ and\
  \bibinfo {author} {\bibfnamefont {G.}~\bibnamefont {Chen}},\ }\href@noop {}
  {\bibfield  {journal} {\bibinfo  {journal} {Nature Comm.},\ }\textbf
  {\bibinfo {volume} {6}},\ \bibinfo {pages} {6755} (\bibinfo {year}
  {2015})}\BibitemShut {NoStop}%
\bibitem [{\citenamefont {Guidry}(1991)}]{guidry91}%
  \BibitemOpen
  \bibfield  {author} {\bibinfo {author} {\bibfnamefont {M.}~\bibnamefont
  {Guidry}},\ }\href@noop {} {\emph {\bibinfo {title} {Gauge Field Theories, an
  introduction with applications}}}\ (\bibinfo  {publisher} {John Wiley {\&}
  Sons, Inc},\ \bibinfo {year} {1991})\BibitemShut {NoStop}%
\bibitem [{\citenamefont {Haug}\ and\ \citenamefont {Jauho}(1996)}]{haug96}%
  \BibitemOpen
  \bibfield  {author} {\bibinfo {author} {\bibfnamefont {H.}~\bibnamefont
  {Haug}}\ and\ \bibinfo {author} {\bibfnamefont {A.-P.}\ \bibnamefont
  {Jauho}},\ }\href@noop {} {\emph {\bibinfo {title} {Quantum Kinetics in
  Transport and Optics of Semiconductors}}}\ (\bibinfo  {publisher}
  {Springer},\ \bibinfo {year} {1996})\BibitemShut {NoStop}%
\bibitem [{\citenamefont {Wang}\ \emph {et~al.}(2008)\citenamefont {Wang},
  \citenamefont {Wang},\ and\ \citenamefont {{L\"u}}}]{wang08review}%
  \BibitemOpen
  \bibfield  {author} {\bibinfo {author} {\bibfnamefont {J.-S.}\ \bibnamefont
  {Wang}}, \bibinfo {author} {\bibfnamefont {J.}~\bibnamefont {Wang}}, \ and\
  \bibinfo {author} {\bibfnamefont {J.~T.}\ \bibnamefont {{L\"u}}},\
  }\href@noop {} {\bibfield  {journal} {\bibinfo  {journal} {Eur. Phys. J. B},\
  }\textbf {\bibinfo {volume} {62}},\ \bibinfo {pages} {381} (\bibinfo {year}
  {2008})}\BibitemShut {NoStop}%
\bibitem [{\citenamefont {Wang}\ \emph {et~al.}(2014)\citenamefont {Wang},
  \citenamefont {Agarwalla}, \citenamefont {Li},\ and\ \citenamefont
  {Thingna}}]{wang14rev}%
  \BibitemOpen
  \bibfield  {author} {\bibinfo {author} {\bibfnamefont {J.-S.}\ \bibnamefont
  {Wang}}, \bibinfo {author} {\bibfnamefont {B.~K.}\ \bibnamefont {Agarwalla}},
  \bibinfo {author} {\bibfnamefont {H.}~\bibnamefont {Li}}, \ and\ \bibinfo
  {author} {\bibfnamefont {J.}~\bibnamefont {Thingna}},\ }\href@noop {}
  {\bibfield  {journal} {\bibinfo  {journal} {Front. Phys.},\ }\textbf
  {\bibinfo {volume} {9}},\ \bibinfo {pages} {673} (\bibinfo {year}
  {2014})}\BibitemShut {NoStop}%
\bibitem [{\citenamefont {Aeberhard}(2011)}]{aeberhard12}%
  \BibitemOpen
  \bibfield  {author} {\bibinfo {author} {\bibfnamefont {U.}~\bibnamefont
  {Aeberhard}},\ }\href@noop {} {\bibfield  {journal} {\bibinfo  {journal} {J.
  Computational Electronics},\ }\textbf {\bibinfo {volume} {10}},\ \bibinfo
  {pages} {394} (\bibinfo {year} {2011})}\BibitemShut {NoStop}%
\bibitem [{\citenamefont {Bruus}\ and\ \citenamefont
  {Flensberg}(2004)}]{bruus04}%
  \BibitemOpen
  \bibfield  {author} {\bibinfo {author} {\bibfnamefont {H.}~\bibnamefont
  {Bruus}}\ and\ \bibinfo {author} {\bibfnamefont {K.}~\bibnamefont
  {Flensberg}},\ }\href@noop {} {\emph {\bibinfo {title} {Many-Body Quantum
  Theory in Condensed Matter Physics, an introduction}}}\ (\bibinfo
  {publisher} {Oxford Univ. Press},\ \bibinfo {year} {2004})\BibitemShut
  {NoStop}%
\bibitem [{\citenamefont {Wang}\ \emph {et~al.}(2007)\citenamefont {Wang},
  \citenamefont {Zeng}, \citenamefont {Wang},\ and\ \citenamefont
  {Gan}}]{jswpre07}%
  \BibitemOpen
  \bibfield  {author} {\bibinfo {author} {\bibfnamefont {J.-S.}\ \bibnamefont
  {Wang}}, \bibinfo {author} {\bibfnamefont {N.}~\bibnamefont {Zeng}}, \bibinfo
  {author} {\bibfnamefont {J.}~\bibnamefont {Wang}}, \ and\ \bibinfo {author}
  {\bibfnamefont {C.-K.}\ \bibnamefont {Gan}},\ }\href@noop {} {\bibfield
  {journal} {\bibinfo  {journal} {Phys. Rev. E},\ }\textbf {\bibinfo {volume}
  {75}},\ \bibinfo {pages} {061128} (\bibinfo {year} {2007})}\BibitemShut
  {NoStop}%
\bibitem [{\citenamefont {Hedin}(1965)}]{hedin65}%
  \BibitemOpen
  \bibfield  {author} {\bibinfo {author} {\bibfnamefont {L.}~\bibnamefont
  {Hedin}},\ }\href@noop {} {\bibfield  {journal} {\bibinfo  {journal} {Phys.
  Rev.},\ }\textbf {\bibinfo {volume} {139}},\ \bibinfo {pages} {A796}
  (\bibinfo {year} {1965})}\BibitemShut {NoStop}%
\bibitem [{\citenamefont {{L\"u}}\ and\ \citenamefont {Wang}(2007)}]{lu2007}%
  \BibitemOpen
  \bibfield  {author} {\bibinfo {author} {\bibfnamefont {J.-T.}\ \bibnamefont
  {{L\"u}}}\ and\ \bibinfo {author} {\bibfnamefont {J.-S.}\ \bibnamefont
  {Wang}},\ }\href@noop {} {\bibfield  {journal} {\bibinfo  {journal} {Phys.
  Rev. B},\ }\textbf {\bibinfo {volume} {76}},\ \bibinfo {pages} {165418}
  (\bibinfo {year} {2007})}\BibitemShut {NoStop}%
\bibitem [{\citenamefont {Zhang}\ \emph {et~al.}(2013)\citenamefont {Zhang},
  \citenamefont {{L\"u}},\ and\ \citenamefont {Wang}}]{zhanglifa2013}%
  \BibitemOpen
  \bibfield  {author} {\bibinfo {author} {\bibfnamefont {L.}~\bibnamefont
  {Zhang}}, \bibinfo {author} {\bibfnamefont {J.-T.}\ \bibnamefont {{L\"u}}}, \
  and\ \bibinfo {author} {\bibfnamefont {J.-S.}\ \bibnamefont {Wang}},\
  }\href@noop {} {\bibfield  {journal} {\bibinfo  {journal} {J. Phys.: Condense
  Matter},\ }\textbf {\bibinfo {volume} {25}},\ \bibinfo {pages} {445801}
  (\bibinfo {year} {2013})}\BibitemShut {NoStop}%
\bibitem [{\citenamefont {Mahan}(2000)}]{mahan00}%
  \BibitemOpen
  \bibfield  {author} {\bibinfo {author} {\bibfnamefont {G.~D.}\ \bibnamefont
  {Mahan}},\ }\href@noop {} {\emph {\bibinfo {title} {Many-Particle
  Physics}}},\ \bibinfo {edition} {3rd}\ ed.\ (\bibinfo  {publisher} {Kluwer
  Academic},\ \bibinfo {year} {2000})\BibitemShut {NoStop}%
\bibitem [{\citenamefont {Biehs}\ \emph {et~al.}(2010)\citenamefont {Biehs},
  \citenamefont {Rousseau},\ and\ \citenamefont {Greffet}}]{Biehs10}%
  \BibitemOpen
  \bibfield  {author} {\bibinfo {author} {\bibfnamefont {S.-A.}\ \bibnamefont
  {Biehs}}, \bibinfo {author} {\bibfnamefont {E.}~\bibnamefont {Rousseau}}, \
  and\ \bibinfo {author} {\bibfnamefont {J.-J.}\ \bibnamefont {Greffet}},\
  }\href@noop {} {\bibfield  {journal} {\bibinfo  {journal} {Phys. Rev.
  Lett.},\ }\textbf {\bibinfo {volume} {105}},\ \bibinfo {pages} {234301}
  (\bibinfo {year} {2010})}\BibitemShut {NoStop}%
\end{thebibliography}%
%%%%%%%%%% Merge with supplemental materials %%%%%%%%%%
%\pagebreak
%\widetext
%\twocolumngrid 
\widetext
\clearpage
\begin{center}
\textbf{\large Supplemental Materials: A microscopic theory for ultra-near-field radiation}
\end{center}
\begin{center}
Jian-Sheng Wang and Jiebin Peng
\end{center}
%%%%%%%%%% Merge with supplemental materials %%%%%%%%%%
%%%%%%%%%% Prefix a "S" to all equations, figures, tables and reset the counter %%%%%%%%%%
\setcounter{equation}{0}
\setcounter{figure}{0}
\setcounter{table}{0}
\setcounter{page}{1}
\makeatletter
\renewcommand{\theequation}{S\arabic{equation}}
\renewcommand{\thefigure}{S\arabic{figure}}
\renewcommand{\bibnumfmt}[1]{[S#1]}
\renewcommand{\citenumfont}[1]{S#1}
%%%%%%%%%% Prefix a "S" to all equations, figures, tables and reset the counter %%%%%%%%%%
%\section{Section 1}
In this supplemental section, we outline the key steps of derivation for the
3D model.  Due to the periodicity of the electron lattice in the transverse directions, the electron Hamiltonian becomes
block diagonal with the Fourier transform of fermion operators defined on lattice sites as
\begin{equation}
c_{l_x,l_y,l_z} = \frac{1}{L} \sum_{{\bf q}_\perp} e^{ i {\bf q}_\perp \cdot 
{\bf l}_\perp a} c({\bf q}_\perp, l_z),
\end{equation}
where the transverse wavevector ${\bf q}_\perp = (q_x, q_y) =  \bigl( \frac{2\pi m}{aL}, \frac{2\pi n}{aL}\bigr)$, $m,n=0, 1, \cdots, L-1$, ${\bf l}_\perp = (l_x, l_y)$, and
$a$, the lattice constant.    Focusing on the right side, letting $c({\bf q}_\perp)$ denote the semi-infinite vector of annihilation operators 
consisting of layers 1, 2, $\cdots$, $l_z$, $\cdots$, the Hamiltonian of the right system and the bath can be written as 
\begin{equation}
H^R_e = \sum_{{\bf q}_\perp} c^\dagger({\bf q}_\perp)
 \left(\matrix{
       v_1 + \epsilon_{2D}({\bf q}_\perp) & -t & 0 & \cdots \cr
                   -t & \epsilon_{2D}({\bf q}_\perp) & -t & 0  \cr
                   0 & -t & \epsilon_{2D}({\bf q}_\perp) & -t \cr
                    \vdots & 0& -t & \ddots 
        }
  \right) c({\bf q}_\perp),
\end{equation}
where $t$ is the hopping parameter and $\epsilon_{2D}({\bf q}_\perp) = 
-2t \bigl( \cos(q_x a) + \cos(q_y a) \bigr)$ is the electron dispersion relation on a two dimensional square lattice.  The free electron Green's function can be found by taking the inverse,
$G_0^r({\bf q}_\perp, E) = \bigl( E + i \eta - H({\bf q}_\perp) \bigr)^{-1}$. $H({\bf q}_\perp)$ is the single particle Hamiltonian displayed as a matrix in the above equation.  For notational simplicity, we'll omit the subscript 0 below.  An explicit expression can be given as
$G^r({\bf q}_\perp, E, l_z, l'_z) = B_{\min(l_z,l'_z)} \lambda^{|l_z-l'_z|}$, where 
$\lambda$ satisfies the quadratic equation
\begin{equation}
t + \bigl(E + i \eta - \epsilon_{2D}({\bf q}_\perp)\bigr) \lambda + t \lambda^2 = 0,
\end{equation}
with modulus $|\lambda|<1$.
We will set an artificial damping to the electrons by choosing $\eta = \hbar/(2\tau)$ where
$\tau$ is the relaxation time related to electron conductivity.  $B_j$ is obtained from a recursion relation, $B_j = -\lambda/t + \lambda^2 B_{j-1}$ with $B_{1} = 
1/\bigl(E + i\eta - \epsilon_{2D}({\bf q}_\perp) - v_1 + t\lambda\bigr)$.

For the vector potential $A^\alpha$, we use exactly the same Fourier transform as for the electrons. This is an approximation in the transverse direction, since the field forms a continuum;  the $z$ direction is still treated as a continuous field.  But
since electrons sit on lattice sites, we only need the values of the field at the lattice sites.  The transverse Fourier transformed version of the photon Green's function is
then,
\begin{equation}
D^{\alpha \beta}({\bf q}_\perp, z,\tau; z', \tau') = 
\sum_{{\bf l}_\perp} D^{\alpha\beta}( {\bf l}_\perp a, z, \tau; {\bf 0}, z', \tau') 
e^{ -i {\bf q}_\perp \cdot 
{\bf l}_\perp a} =  
-\frac{i}{\hbar} \bigl\langle T_\tau A^\alpha({\bf q}_\perp, z, \tau)
A^\beta(-{\bf q}_\perp, z', \tau') \bigr\rangle.
\end{equation}
To obtain the self-energies of the photons, we expand the exponential term containing the vector potential in Eq.~(\ref{eqegint}) of the main text to second order in $A^\alpha$. The linear term gives the usual current-vector potential interaction, producing an associated self-energy in contour time as, after the standard diagrammatic analysis:
\begin{equation}
\Pi^{\alpha\beta}_{(1)}({\bf q}_\perp, l_z,\tau; l'_z, \tau') = - \frac{i}{L^2\hbar}
\bigl\langle T_\tau I^\alpha({\bf q}_\perp, l_z, \tau)
I^\beta(-{\bf q}_\perp, l'_z, \tau') \bigr\rangle,
\end{equation}
where the transverse components of the current operators, $\alpha = x$, $y$, are given by 
\begin{equation}
I^\alpha({\bf q}_\perp, l_z, \tau) = (-e) \sum_{{\bf p}_\perp, {\bf p}'_\perp}
v(p_\alpha, p'_\alpha) c^\dagger({\bf p}_\perp, l_z, \tau)
c({\bf p}'_\perp, l_z, \tau) \delta({\bf p}'_\perp - {\bf p}_\perp - {\bf q}_\perp).  
\end{equation}
The `velocity' of the electron is $v(p_\alpha, p'_\alpha) = \frac{a t}{\hbar} \bigl(
\sin(p_\alpha a) + \sin(p'_\alpha a) \bigr)$, and the last term $\delta$ is the Kronecker delta.
The $z$ component is different, given by 
\begin{equation}
I^z({\bf q}_\perp, l_z, \tau) = { i eat \over 2 \hbar} \sum_{{\bf p}_\perp, {\bf p}'_\perp}
\Bigl[ c^\dagger({\bf p}_\perp, l_z, \tau) \Delta c({\bf p}'_\perp, l_z, \tau)
 -\Delta c^\dagger({\bf p}_\perp, l_z, \tau) c({\bf p}'_\perp, l_z, \tau)
\Bigr]\delta({\bf p}'_\perp - {\bf p}_\perp - {\bf q}_\perp),  
\end{equation}
where we have defined a new operator, the central difference, 
$\Delta c({\bf p}_\perp, l_z, \tau) = c({\bf p}_\perp, l_z+1, \tau) - c({\bf p}_\perp, l_z-1, \tau)$, and similarly for $\Delta c^\dagger({\bf p}_\perp, l_z, \tau) $. 

The self-energies can be expressed in terms of the electron Green's function, $G$, by applying the Wick theorem.  The transverse sector, $\alpha,\beta=x,y$, is
\begin{equation}
\Pi^{\alpha\beta}_{(1)}({\bf q}_\perp, l_z,\tau; l'_z, \tau') = - \frac{i\hbar e^2}{L^2}
\sum_{{\bf p}^1_\perp, {\bf p}^2_\perp}
v(p^1_\alpha, p^2_\alpha) 
v(p^1_\beta, p^2_\beta) 
G({\bf p}^1_\perp, l_z, \tau; l'_z, \tau') 
G({\bf p}^2_\perp, l'_z, \tau'; l_z, \tau) 
\delta({\bf p}^1_\perp - {\bf p}^2_\perp - {\bf q}_\perp). 
\end{equation}
And the $\alpha z$ sector is 
\begin{eqnarray}
\Pi^{\alpha z}_{(1)}({\bf q}_\perp, l_z,\tau; l'_z, \tau') &=&  \frac{e^2at}{2 L^2}
\sum_{{\bf p}^1_\perp, {\bf p}^2_\perp}
v(p^1_\alpha, p^2_\alpha) 
\Bigl[ 
G_{c\, \Delta c^\dagger}({\bf p}^1_\perp, l_z, \tau; l'_z, \tau') 
G ({\bf p}^2_\perp, l'_z, \tau'; l_z, \tau)
\nonumber \\
&& \qquad -\,  
G({\bf p}^1_\perp, l_z, \tau; l'_z, \tau') 
G_{\Delta c\, c^\dagger} ({\bf p}^2_\perp, l'_z, \tau'; l_z, \tau)
\Bigr]
\delta({\bf p}^1_\perp - {\bf p}^2_\perp - {\bf q}_\perp),
\end{eqnarray}
and similarly
\begin{eqnarray}
\Pi^{z\alpha}_{(1)}({\bf q}_\perp, l_z,\tau; l'_z, \tau') &=&  \frac{e^2at}{2 L^2}
\sum_{{\bf p}^1_\perp, {\bf p}^2_\perp}
v(p^1_\alpha, p^2_\alpha) 
\Bigl[ 
G({\bf p}^1_\perp, l_z, \tau; l'_z, \tau') 
G_{c \, \Delta c^\dagger} ({\bf p}^2_\perp, l'_z, \tau'; l_z, \tau)
\nonumber \\
&& \qquad -\,  
G_{\Delta c c^\dagger}({\bf p}^1_\perp, l_z, \tau; l'_z, \tau') 
G ({\bf p}^2_\perp, l'_z, \tau'; l_z, \tau) 
\Bigr]
\delta({\bf p}^1_\perp - {\bf p}^2_\perp - {\bf q}_\perp). 
\end{eqnarray}
We have introduced a self-explanatory notation where $G_{AB}(\tau, \tau') = 
-\frac{i}{\hbar} \bigl\langle T_\tau A(\tau) B(\tau') \bigr\rangle$.   The $zz$ component is 
\begin{eqnarray}
\Pi^{zz}_{(1)}({\bf q}_\perp, l_z,\tau; l'_z, \tau') &=& \frac{i (eat)^2}{4\hbar L^2}
\sum_{{\bf p}^1_\perp, {\bf p}^2_\perp}
\Bigl[ G_{\Delta c\, c^\dagger}({\bf p}^1_\perp, l_z, \tau; l'_z, \tau') 
G_{\Delta c\, c^\dagger}({\bf p}^2_\perp, l'_z, \tau'; l_z, \tau) \nonumber \\
&&  - \, G({\bf p}^1_\perp, l_z, \tau; l'_z, \tau') 
G_{\Delta c \, \Delta c^\dagger} ({\bf p}^2_\perp, l'_z, \tau'; l_z, \tau) 
 -  G_{\Delta c \, \Delta c^\dagger}({\bf p}^1_\perp, l_z, \tau; l'_z, \tau') 
G ({\bf p}^2_\perp, l'_z, \tau'; l_z, \tau) \nonumber \\
&& \quad +\, G_{c\, \Delta c^\dagger}({\bf p}^1_\perp, l_z, \tau; l'_z, \tau') 
G_{c\, \Delta c^\dagger}({\bf p}^2_\perp, l'_z, \tau'; l_z, \tau)
\Bigr]
\delta({\bf p}^1_\perp - {\bf p}^2_\perp - {\bf q}_\perp). 
\end{eqnarray}
The contour time functions can be transformed into the energy (or frequency) domain analogously to the formulas given in the main texts.

The quadratic term $(A^\alpha)^2$ in the expansion gives a plasmon contribution,
which is diagonal in spin and site indices,  in energy space ($E = \hbar \omega$)
\begin{equation}
\Pi^{\alpha\beta}_{(2)}( {\bf q}_\perp, E, l_z, l'_z) = 
- \frac{i \hbar e^2}{mL^2} \delta_{\alpha,\beta} \delta_{l_z, l'_z} \sum_{{\bf p}_\perp}
\int_{-\infty}^{+\infty} \frac{dE'}{2 \pi \hbar } \cos(p_\alpha a) G^<({\bf p}_\perp,
E', l_z, l_z),\qquad \alpha,\beta = x,y.
\end{equation}
The mass is defined by $t = \hbar^2 /(2 m a^2)$.
The $zz$ component is 
\begin{eqnarray}
\Pi^{zz}_{(2)}( {\bf q}_\perp, E, l_z, l'_z) &=& 
- \frac{i \hbar e^2}{4mL^2}  \delta_{l_z, l'_z} \sum_{{\bf p}_\perp}
\int_{-\infty}^{+\infty} \frac{dE'}{2 \pi \hbar }  \Bigl[ 
G^<({\bf p}_\perp,E', l_z, l_z+1) + 
G^<({\bf p}_\perp,E', l_z+1, l_z) + \nonumber \\ 
&& \qquad \qquad\qquad \qquad G^<({\bf p}_\perp,E', l_z, l_z-1) + 
G^<({\bf p}_\perp,E', l_z-1, l_z)
\Bigr].
\end{eqnarray}
The total self-energy is $\Pi = \Pi_{(1)} +\Pi_{(2)}$.  In actual calculations, we have
taken ${\bf q}_\perp = {\bf 0}$ in the photon self-energy expression.  This is a valid approximation since the thermal wavelengths of the photons are much longer than 
that of the electrons.

To solve the Dyson equation, we need an analytic expression for the free
photon Green's function $D_0^r$.  This can be obtained from the second quantization representation of the vector potential [S1],
\begin{equation}
{\bf A}({\bf r}, t) = 
\sum_{{\bf q}, \sigma=1,2} \sqrt{\frac{\hbar}{2 \epsilon_0 \omega_{\bf q} a^3 L^3}}
\; {\bf e}({\bf q}, \sigma) \left( a_{{\bf q}, \sigma} e^{i ( {\bf q} \cdot {\bf r} -  
\omega_{\bf q} t)} + {\rm .h.c.} \right),
\end{equation}
where ${\bf e}({\bf q}, 1)$ and  ${\bf e}({\bf q}, 2)$ are the two unit polarization vectors perpendicular to $\bf q$.   Taking into account the fact that we already have 
${\bf q}_\perp$, and following the definition of the retarded Green's function, we find
\begin{equation}
D^{r,\alpha\beta}_0({\bf q}_\perp, \omega, z, z')  = \int_{-\infty}^{+\infty}\!\! {dq_z \over 2\pi} \,
{ \left( \delta_{\alpha\beta} - \frac{q_\alpha q_\beta}{q^2}\right) 
e^{i q_z( z - z')} \over a^2 \epsilon_0 \Bigl( (\omega + i0^+) ^2 - c^2 q_\perp^2
- c^2 q_z^2\Bigr) },   
\end{equation}
where $q^2 = | {\bf q} |^2 = q_\perp^2 + q_z^2$.  This integral can be performed using the residue theorem.  We obtain
\begin{eqnarray}
D_0^{r,\alpha\beta} &=& \delta_{\alpha\beta}\, d - q_\alpha q_\beta F, \quad \alpha, \beta = x,y\\
D_0^{r,\alpha z} &=& D_0^{r,z\alpha} = {\rm sgn}(z-z')q_\alpha ( B - A)/C, \\
D_0^{r,zz} &=& q_\perp^2 F.
\end{eqnarray}
We have introduced the shorthand notations $A=e^{i \tilde{q}_z |z-z'|}$,
$B=e^{-q_\perp|z-z'|}$,  $d = A/(a^2 \epsilon_0 2 i c^2 \tilde{q}_z)$,
$F= (A/\tilde{q}_z + i B/q_\perp)/C$, $C=a^2\epsilon_0 2 i \omega^2$, and $\tilde{q}_z = \pm \sqrt{ 
[(\omega + i0^+)/c]^2 - q_\perp^2}$, where the sign is chosen such that
${\rm Im}\, \tilde{q}_z > 0$.
 
Finally, the Poynting vector average is computed from
\begin{equation}
\bigl\langle S^z(z) \bigr\rangle = \frac{1}{\mu_0 L^2} \sum_{{\bf q}_\perp} 
\int_0^\infty { d\omega \over \pi}  \hbar \omega\, {\rm Re} \,\left( - {\partial \over \partial z'} \sum_{\gamma=x,y} D^{<,\gamma\gamma}({\bf q}_\perp, \omega, z,z')\Big|_{z'=z} \right),
\end{equation}
where $\mu_0 = 1/(c^2 \epsilon_0)$ is the vacuum permeability.  $D^<$ is further 
expressed in terms of the retarded Green's function through the Keldysh equation. 
We do not include the photon bath terms --- this is equivalent to setting the bath temperatures to 0 ---
and $\Pi^<$ follows the fluctuation-dissipation theorem,
under the Born approximation, since $G_0$ is in equilibrium. 

\vskip 12pt

\noindent [S1] C. Cohen-Tannoudji, J. Dupont-Roc, and G. Grynberg,
{\sl Photons \& Atoms, introduction to quantum electrodynamics}, Chap. III.
Wiley-VCH (2004).

\end{document}